\documentclass[12pt,a4paper]{article}

\usepackage{amsmath,amssymb}
\usepackage[nosort]{cite}
\usepackage[bulletsep]{collref}
\usepackage{feynmp}
\usepackage{multirow}
\usepackage{listings}
\usepackage{longtable}
\usepackage{multicol}
\usepackage{hhline}

\usepackage{epsfig,psfrag}

\newcommand{\cO}{{\cal O}}
\newcommand{\cN}{{\cal N}}
\def \Tr {\mathop{\rm Tr}\nolimits}
\newcommand{\ft}[2]{{\textstyle\frac{#1}{#2}}}

\newcommand \vev [1] {\langle{#1}\rangle}

\begin{document}

\begin{flushright}
HU-EP-11/01\\
\end{flushright}
%\vspace{25mm}
\vspace{3mm}

\begin{center}
{\Large\bf\sf  
%Scattering Amplitudes in Gauge Theories
%\\
Dual conformal symmetry at loop level; massive regularization\footnote{Invited review for a special issue of Journal of Physics A devoted to
 ``Scattering Amplitudes in Gauge Theories'', R. Roiban(ed), M. Spradlin(ed),
A. Volovich(ed).}
}

\vskip 5mm 
Johannes M. Henn$^{a}$\\
{\tt %\phantom{aaa}
henn@physik.hu-berlin.de}
\end{center}

\begin{center}
$^{a}${\em Institut f\"ur Physik\\
Humboldt-Universit\"at zu Berlin, \\
Newtonstra\ss{}e15, D-12489 Berlin, Germany}
\end{center}

\vskip 2mm

\begin{abstract}
Dual conformal symmetry has had a huge impact on our understanding
of planar scattering amplitudes in $\cN=4$ super Yang-Mills.
At tree level, it combines with the original conformal symmetry generators
to a Yangian algebra, a hallmark of integrability, and helps in determining the
tree-level amplitudes. The latter are now known in closed form.
At loop level, it determines the functional form of the four- and
five-point scattering amplitudes to all orders in the coupling constant, and gives
restrictions at six points and beyond.
The symmetry is best understood at loop level in terms of a novel AdS-inspired
infrared regularization which makes the symmetry exact, despite the infrared divergences. 
This has important consequences for the basis of loop integrals in this theory.
Recently, a number of selective 
reviews have appeared which discuss dual conformal symmetry, 
mostly at tree level. Here, we give an up-to-date account
of dual conformal symmetry, focussing on its status at loop level.
\end{abstract}

\vfil\break

\section{Introduction}

The last years have seen exciting progress in the understanding of scattering amplitudes
in gauge theories, in particular in the maximally supersymmetric $\cN=4$ super Yang-Mills (SYM) theory
in the planar limit.
Many of these developments in the field theory have been driven by the discovery of new symmetries, as
well as by exploiting the analytic properties of scattering amplitudes, including their infrared structure.
Apart from providing us with exciting new results in $\cN=4$ SYM, 
these advances allow us to gain insights into the structure of loop amplitudes in general
and also have applications for theories with less or no supersymmetry.
This review is part of the volume \cite{JPAvolume} that aims to give an up-to-date account of these developments.
\\

This review is organized as follows: We begin by motivating the use of dual coordinates for planar graphs
and by showing hints for a dual conformal symmetry of loop integrals contributing to scattering amplitudes in $\cN=4$ SYM in section \ref{sect-historical}.
The symmetry is obscured in part by the presence of infrared divergences.
In section \ref{sect-today}, we introduce an infrared regulator that is motivated by the AdS/CFT correspondence and that
allows to make dual conformal symmetry exact at loop level. We discuss various features of this setup and its
implications for the loop-level integral basis. We comment on recent developments for computing loop integrands
using recursion relations.
In section \ref{sect-integrals} we present aspects of loop integrals and their analytical computation, with a focus on the infrared regularization of section \ref{sect-today}.
We also give an example of an integral belonging to a special class of dual conformal integrals with certain numerator factors that are relevant for $\cN=4$ SYM
and satisfy simple differential equations.
We motivate a possible connection between the differential equations and the conformal symmetry of $\cN=4$ SYM by giving an 
example of a Yangian invariant integral.\\

\section{Hints for dual conformal symmetry\protect\footnote{This section is organized 
in a rather historical fashion for pedagogical purposes.
A better understanding of dual conformal symmetry is now available in terms of the mass regulator
that we discuss in section \ref{sect-today}.}
}
\label{sect-historical}

\begin{figure}[t]
\psfrag{a}[cc][cc]{(a)} 
\psfrag{b}[cc][cc]{(b)} 
\psfrag{c}[cc][cc]{(c)} 
\psfrag{d}[cc][cc]{(d)} 
\psfrag{p1}[cc][cc]{$p_{1}$}
\psfrag{p2}[cc][cc]{$p_{2}$}
\psfrag{p3}[cc][cc]{$p_{3}$}
\psfrag{p4}[cc][cc]{$p_{4}$}
 \centerline{
 {\epsfxsize14cm  \epsfbox{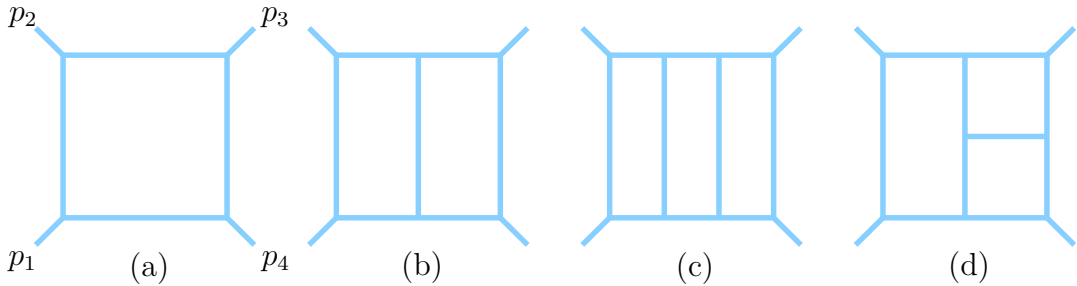}}
} 
\caption{\small
Loop integrals appearing up to three loops in the four-point amplitude.
Numerator factors independent of the loop momentum and a loop-dependent numerator
in diagram (d) are not displayed.
}\label{fig-loop}
\end{figure}

The first hints for a dual conformal symmetry of scattering amplitudes in planar $\cN=4$ SYM came from
inspecting the loop integrals contributing to the four-gluon amplitudes \cite{Drummond:2006rz} 
(see also \cite{Lipatov:1993qn}.)
To three loops and up to a trivial tree-level factor, they are given by a linear combination 
of the integrals shown in Figure \ref{fig-loop} \cite{Anastasiou:2003kj,Bern:2005iz}. 
Although these integrals superficially look like diagrams obtained from a $\phi^3$ theory, one should
keep in mind that, at least in principle, they are the result of summing over a large number of 
Feynman diagrams. In practice, one often uses methods that are based
on the analytic properties of the perturbative S-matrix \cite{Bern:1994zx,Bern:1994cg,Britto:2004ap,Britto:2005fq}
(see also \cite{JPAunitarity1, JPAunitarity2} of this volume)
and that do not make explicit use of Feynman diagrams.\\

\begin{figure}[t]
\psfrag{x1}[cc][cc]{$x_{1}$}
\psfrag{x2}[cc][cc]{$x_{2}$}
\psfrag{x3}[cc][cc]{$x_{3}$}
\psfrag{x4}[cc][cc]{$x_{4}$}
\psfrag{xj}[cc][cc]{$x_{0}$}
\psfrag{a}[cc][cc]{(a)} 
\psfrag{b}[cc][cc]{(b)} 
\psfrag{c}[cc][cc]{(c)} 
\psfrag{d}[cc][cc]{(d)} 
 \centerline{
 {\epsfxsize14cm  \epsfbox{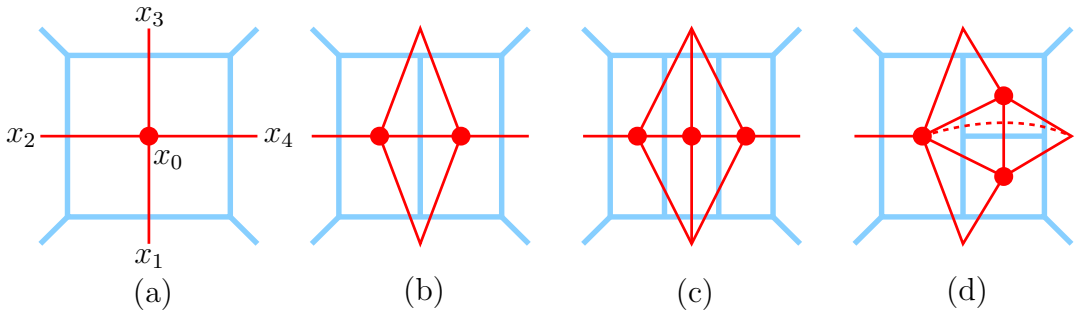}}
} 
\caption{\small
Dual representation of the integrals of Fig. \ref{fig-loop}. Vertices denote dual integration
points. The dashed line in Fig. (d) denotes an internal numerator factor. The latter is
required by dual conformal symmetry, as explained in the text.
}\label{fig-dual}
\end{figure}

The fact that only few integral topologies remain at the end is very remarkable.
It was understood over the last years that it is the consequence of a new symmetry of 
planar scattering amplitudes in $\cN=4$ SYM, as we discuss presently. 
As an explicit example, at the one-loop level and for four points, only the 
scalar box integral shown in Figure \ref{fig-loop}(a) appears.
It is given by
\begin{equation}\label{dimregbox}
I^{(1) \epsilon} = \int \frac{d^{D}k}{i \pi^{D/2}} \frac{ (p_{1} + p_{2})^2 (p_{2}+p_{3})^2 }{k^2 (k+p_{1})^2 (k+p_{1}+p_{2})^2 (k-p_{4})^2} \,,
\end{equation}
with the on-shell conditions $p_{i}^2 = 0$, and where the calculation leading to (\ref{dimregbox})
has been done in dimensional regularization, with $D=4-2\epsilon$ and $\epsilon<0$ to regularize infrared divergences.
For a generic theory, also triangle integrals could have appeared (bubble integrals would be UV divergent and
are therefore excluded in a UV-finite theory.)
Being a planar integral, we can unambiguously define
dual or region coordinates $x_{i}$ by
\begin{equation}\label{dualcoordinates}
p_{i}^{\mu} = x_{i}^{\mu} -x_{i+1}^{\mu} \,,
\end{equation}
with the cyclicity condition $x_{i+4} \equiv x_{i}$. 
The on-shell conditions become $x_{i,i+1}^2 = 0$.
For the one-loop box integral (\ref{dimregbox}) this leads to
\begin{align}\label{I4dual}
I^{(1) \epsilon} = \int \frac{d^{D}x_{0}}{i \pi^{D/2}} \frac{ x_{13}^2 x_{24}^2 }{x_{01}^2  x_{02}^2 x_{03}^2 x_{04}^2}\,.
\end{align}
Here the change of variables $k^\mu = x_{0}^\mu - x_{1}^\mu $ was done, and the resulting 
dual graph is shown in Fig. \ref{fig-dual}(a). See \cite{Nakanishi} for a reference on
graph theory discussing dual graphs.
The use of dual variables for planar integrals is in fact very useful, independently of the symmetry that we are going to discuss.
For example, imagine we wish to write down the Feynman parametrization for a generic one-loop diagram.
Then, if $\alpha_{i}$ is the Feynman parameter associated to the propagator $1/x_{0i}^2$,
the argument of the denominator appearing in the Feynman parameter integral is simply $\sum_{i<j} x_{ij}^2 \alpha_{i} \alpha_{j}$, see e.g. \cite{smirnov2006feynman}.
\\

Written in this form, (\ref{I4dual}) is reminiscent of integrals appearing in the study of
position space correlation functions of protected operators, i.e. operators with 
zero anomalous dimension, in $\cN=4$ SYM, see e.g. \cite{Eden:2000mv}.
The difference is that in those correlation functions, the $x_{i}^{\mu}$ 
are unconstrained variables (i.e. they do not satisfy the on-shell
conditions $x_{i,i+1}^2 = 0$) and that the integration measure is four-dimensional.
In that case the integrals have an SO(4,2) conformal symmetry. While Poincar\'{e}
symmetry is manifest, invariance under special conformal transformations
can be best seen by considering inversions.
The transformations are
\begin{align}\label{dualinversions}
x_{i}^{\mu} \to \frac{x_{i}^{\mu}}{x_{i}^{2}} \,,\qquad x_{ij}^2 \to \frac{x_{ij}^2}{x_{i}^2 x_{j}^2} \,,\qquad d^{D}x_{0} = d^{D}x_{0} (x_{0}^{2})^{-D} \,.
\end{align}
We see that for $D=4$, all factors of $x^2 $ from (\ref{dualinversions}) cancel precisely in (\ref{I4dual}), and the integral is indeed (dual) conformal invariant.
The dual conformal symmetry of the off-shell ladder integrals was first
noted by Broadhurst \cite{Broadhurst:1993ib} and used to explain the equivalence of 
three- and four-point ladder integrals \cite{Usyukina:1993ch}, which are related by conformal
transformations.\\

Coming back to the scattering amplitudes, we recall that we have (\ref{I4dual}) with $D=4-2 \epsilon$
and $\epsilon < 0$. One cannot set $D=4$ because of infrared divergences.
Therefore the symmetry of this integral is only approximate.
We will see in section \ref{sect-today} how this problem can be cured, but for
the moment let us discuss the symmetry in this naive sense.
The crucial observation made in \cite{Drummond:2006rz} is that this integral and
all other integrals contributing up to three loops to the four-gluon amplitude,
which were obtained in the pioneering work of \cite{Anastasiou:2003kj,Bern:2005iz},
are invariant (naively) under conformal transformations in the dual space of
the $x_{i}^{\mu}$ variables.
Integrals having this property are sometimes called ``pseudoconformal''.\\

The dual diagrams of the four-point integrals up to three loops 
are shown in Fig. \ref{fig-dual}, and it is easy to see that they all
have the above property: what one needs to check is that
the conformal weight of each dual integration point is cancelled
by propagator and numerator factors attached to it. 
If there is no numerator, this means that exactly four propagators
need to be attached to each integration point, which is the case
for the integrals shown in Fig. \ref{fig-dual}(a)-(c).
We remark that triangle subgraphs are forbidden by dual
conformal symmetry. Fig. \ref{fig-dual}(d) is an example
of a dual conformal integral with non-trivial, i.e. loop-dependent,
numerator factor. The latter is indicated by a dashed line and is
in fact required to cancel the conformal weight at the integration
point that is joined by five propagators.\\

Dual conformal symmetry seems to be a property of planar scattering amplitudes 
in $\cN=4$ SYM and its presence was
confirmed at higher loop orders \cite{Bern:2006ew,Drummond:2007aua} as well.  
It is also a useful guiding principle for finding the correct loop integrands
for amplitudes at higher loops or with more external legs \cite{Bern:2007ct,Bern:2008ap,Kosower:2010yk}.
This is of great practical help, especially when computations are done
employing (generalized) unitarity. If the basis of loop integrals is known, unitarity
cuts can be used to determine the (rational) coefficients of the integrals.
We remark that although presently dual conformal symmetry applies to
planar amplitudes only, its existence can also be useful for non-planar studies
thanks to relations between planar and non-planar amplitudes, see e.g. \cite{Bern:2010ue,Bern:2010tq}.\\

In fact from the discussion above it is easy to find rules for writing
down dual conformal integrals. An important restriction comes from 
the fact that loop integrands have the structure ``numerator $\times$ propagators'',
where by propagators we mean products of factors like $1/p^2$ \footnote{Since 
scattering amplitudes are gauge invariant we can assume the Feynman gauge
for this discussion. In other gauges the propagator denominators could be
more complicated, or higher powers of $p^2$ could appear.}.
We have already seen that each dual integration point has to be joined
by at least four propagators. If there are more than four propagators 
joining it, the excess in conformal weight has to be cancelled
by appropriate numerator factors. The latter can be inverse
propagators as in Fig.~\ref{fig-dual}(d), or in general also suitably defined traces built from dual variables.\\

The above considerations are very helpful for restricting the loop {\it integrand} of
scattering amplitudes.
In order to make quantitative predictions about the functions obtained {\it after integration}, it
is important to understand the breaking of the symmetry near four dimensions.
Hints for how to do this came from the AdS/CFT correspondence, which suggests a
surprising relation between scattering amplitudes and certain light-like Wilson loops \cite{Alday:2007hr,Drummond:2007aua,Brandhuber:2007yx,Drummond:2007cf}.
This conjectured duality is reviewed in \cite{Alday:2008yw,Henn:2009bd},
and in \cite{JPAspecialWLduality} of this volume.
The light-like Wilson loops appearing in the duality are defined in coordinate
space. They have $n$ cusps which lie precisely at the positions indicated by the dual coordinates 
of equation (\ref{dualcoordinates}). 
The dual conformal symmetry of the scattering amplitudes is then identified with the
conventional conformal symmetry of the Wilson loops.
Importantly, the breaking of the latter is
controlled to all orders in the coupling constant by anomalous Ward identities. 
Admitting the duality with the
(maximally-helicity-violating) scattering amplitudes, the Ward identities can be applied to
the latter.
Let us now quote the form of the Ward identities. We use $M_{n}$ to denote the color-ordered MHV amplitude, with the tree-level
term factored out. The universal form of infrared divergences suggests to write $\log M_{n} = D_{n} + F_{n} + \cO(\epsilon)$
as the sum of a divergent term $D_{n}$, a finite term $F_{n}$, and $\cO(\epsilon)$ corrections.
Given the universal form of $D_{n}$, the Ward identities can be written for $F_{n}$ as \cite{Drummond:2007cf,Drummond:2007au}
\begin{align}\label{dcs-ward}
K^{\mu} F_{n} =  \frac{1}{2} \Gamma_{\rm cusp}(a) \sum_{i=1}^{n} \left[ x_{i,i+1}^{\mu} \log \frac{x_{i,i+2}^2}{x_{i-1,i+1}^2} \right]  \,,
\end{align}
where
\begin{align}\label{kboost}
K^{\mu} = \sum_{i=1}^{n} \left[ 2 x_{i}^\mu x_{j}^\mu \frac{\partial}{\partial x^\nu_{i}} - x_{i}^2 \frac{\partial}{\partial x_{i \mu}} \right] \,,
\end{align}
is the generator of conformal boosts.
The cusp anomalous dimension $\Gamma_{\rm cusp}$ \cite{Korchemskaya:1992je} is conjectured to be governed to all loop orders
by an integral equation \cite{Beisert:2006ez}.
At four- and five-points, equation (\ref{dcs-ward}) has a unique solution (to all orders in the coupling constant), which coincides with the
Bern-Dixon-Smirnov ansatz \cite{Anastasiou:2003kj,Bern:2005iz} and agrees with the AdS calculation of \cite{Alday:2007hr} for four points.
Starting from six points, (\ref{dcs-ward}) determines $F_{n}$ up to an a priori arbitrary (coupling-dependent) function of dual conformal
cross-rations \cite{Drummond:2007cf,Drummond:2007au}, called the remainder function.\\

It is important to stress that this Ward identity applies to (the logarithm of) an
amplitude, not to individual integrals. The reason is that the infrared divergences,
which are responsible for the anomaly on the r.h.s. of (\ref{dcs-ward}), take a simple form only 
for that quantity.
Since infrared divergences are universal it is natural to expect that a generalization
to non-MHV amplitudes holds as well \cite{Drummond:2008vq}. 
This required the generalization to a dual
superconformal space \cite{Drummond:2008vq},
which is reviewed in \cite{JPAspecialdualsuperconf} of this volume.
These predictions were confirmed in various cases,
at tree-level \cite{Drummond:2008vq,Brandhuber:2008pf,Drummond:2008cr},
one loop \cite{Drummond:2008bq,Elvang:2009ya,Brandhuber:2009kh},
and numerically for the six-point NMHV two-loop amplitude \cite{Kosower:2010yk}.
\\

In the above dimensional regularization was used to regulate the IR divergences
of the scattering amplitudes. In fact, this regulator is not best suited
for exploiting the dual conformal symmetry. Although the dual conformal
anomaly of equation (\ref{dcs-ward}) is very simple, 
the action of $K^{\mu}$ on a generic 
``pseudoconformal'' integral is in general very complicated.
This makes the notion of ``pseudoconformal''
integrals rather vague, and in fact it is hard to give a mathematically
concise definition for them (one might think that this can be cured by going off-shell, 
but that can lead to other problems, such as loss of gauge invariance.)
In the next section, we will introduce an alternative regulator which allows to
realize dual conformal symmetry at loop level without an anomaly, and
thereby can be used to make the notion of a dual conformally invariant integrals precise.
This is obviously of great importance in the context of the loop integral basis
that was alluded to earlier.\\

\section{Scattering amplitudes on the Coulomb branch}
\label{sect-today}

In the previous section we saw that the necessity
to regulate the scattering amplitudes obscured the 
dual conformal symmetry, and in particular the dimensional infrared regulator breaks
the latter. As we will review presently, it is possible to regulate the infrared divergences in a 
different way that allows to preserve dual conformal symmetry at loop level \cite{Alday:2009zm}.
We will first explain how to regulate the IR divergences by introducing certain Higgs masses,
and then discuss how the symmetry manifests itself.
\\

\begin{figure}
\psfrag{ads5}[cc][cc]{AdS${}_5$}
\psfrag{N}[ll][ll]{$N$ D3-branes}
\psfrag{M}[ll][ll]{$M$ D3-branes}
\psfrag{z0}[rr][rr]{$z=0$}
\psfrag{zi}[rr][rr]{$z_{i}=1/m_{i}$}
\psfrag{aba}[cc][cc]{(a)}
\psfrag{p1}[cc][cc]{$p_{2}$}
\psfrag{p2}[cc][cc]{$p_{3}$}
\psfrag{p3}[cc][cc]{$p_{4}$}
\psfrag{p4}[cc][cc]{$p_{1}$}
\psfrag{i1}[cc][cc]{\tiny $i_{2}$}
\psfrag{i2}[cc][cc]{\tiny $i_{3}$}
\psfrag{i3}[cc][cc]{\tiny$i_{4}$}
\psfrag{i4}[cc][cc]{\tiny$i_{1}$}
\psfrag{n}[cc][cc]{\tiny$j$}
\psfrag{m}[cc][cc]{\tiny$k$}
\psfrag{box2aA}[cc][cc]{(b)}
 \centerline{
{\epsfysize6cm
\epsfbox{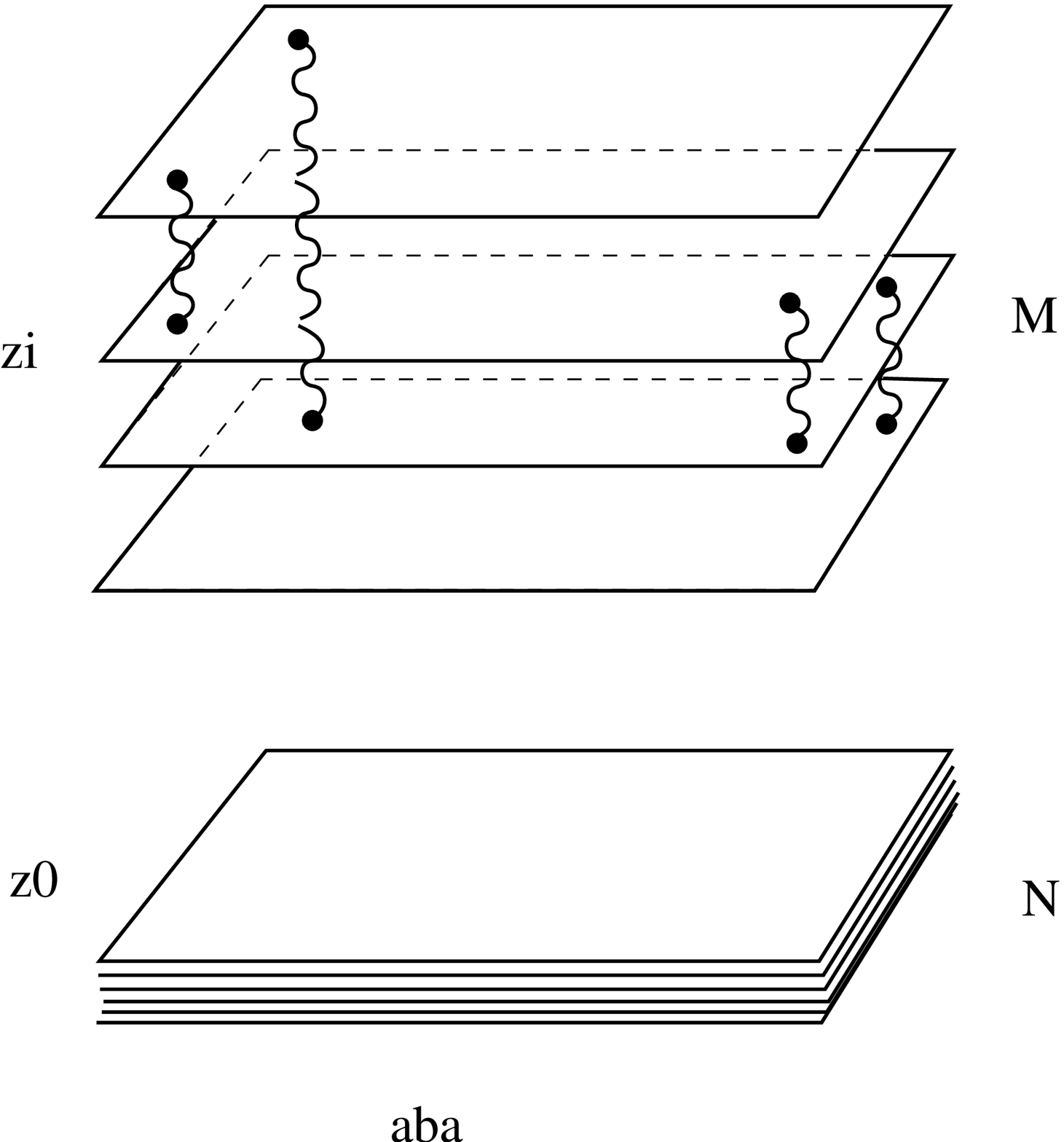}}
\hspace{3cm}
{\epsfysize5cm
\epsfbox{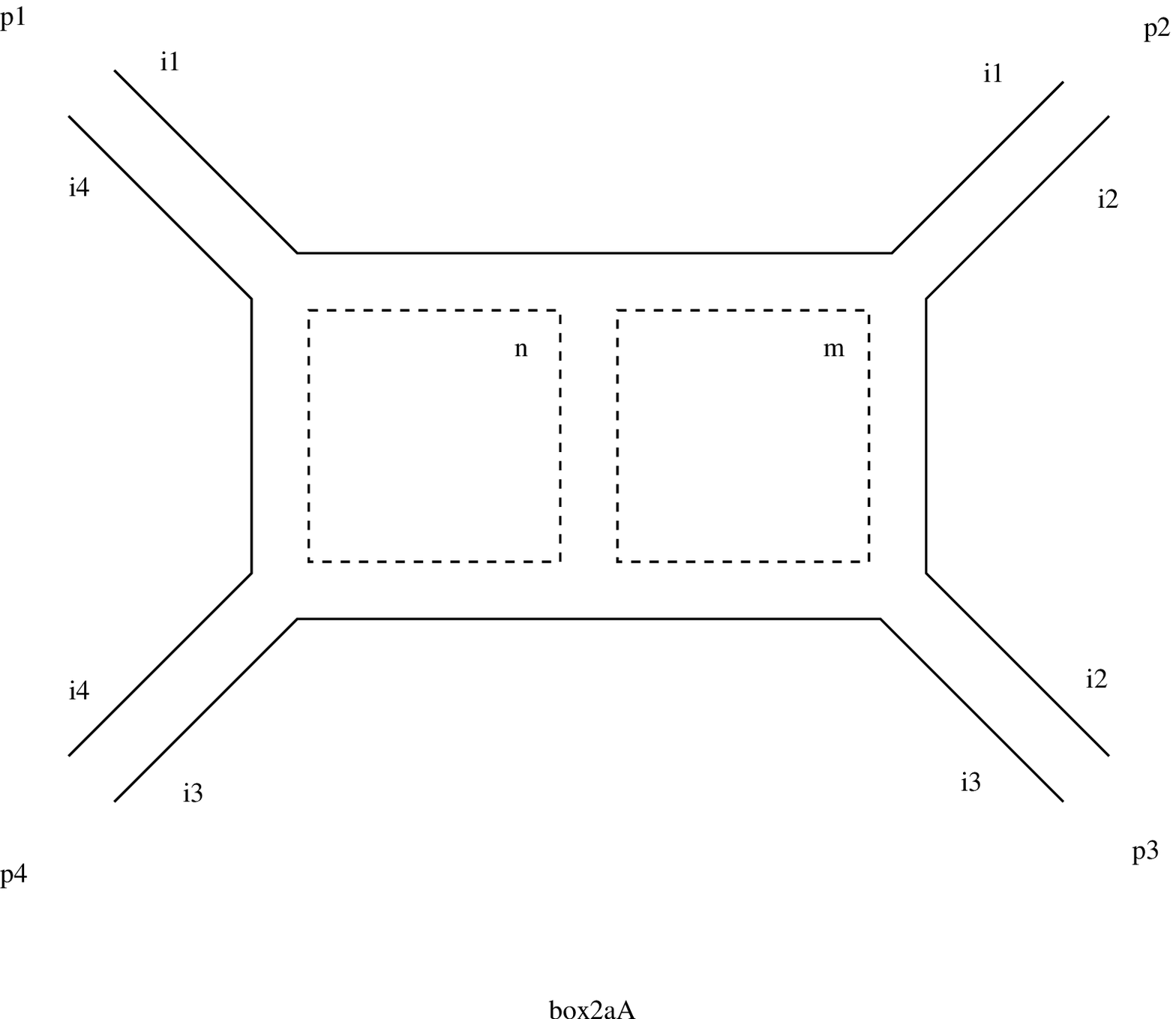}}
\vspace{0.5cm}}
\caption{\small
(a) String theory description for the scattering of $M$ gluons in the large $N$ limit.
Putting the $M$ D3-branes at different positions $z_{i}\neq 0$ serves as a regulator
and also allows to exhibit dual conformal symmetry. (b) Gauge theory analogue of (a): 
a sample two-loop integral at large $N$, in double line notation. Mixed full/dashed lines 
correspond to massive propagators.
Picture from \cite{Alday:2009zm}.}
\label{fone}
\end{figure}

The idea is to start with a gauge group $U(N+M)$ and to break it to $U(N) \times U(M)$.
Let the fields associated to the broken part of the gauge group have mass $m$.
If one scatters $U(M)$ fields and takes $N \gg M$, the dominant diagrams 
are those where a massive particle runs on the perimeter of the diagrams, and
the interior is massless, see Fig.~\ref{fone}(b). The masses on the perimeter regulate the infrared divergences.
In the limit $m \to 0$, we approach the four-dimensional massless theory. Compared
to dimensional regularization, the IR divergences then manifest themselves as $\log^{i} m^2 $
as opposed to $\epsilon^{-i}$, with $i \le 2L$ and $L$ being the loop order.\\

There are a number of technical advantages associated with this fact \cite{Alday:2009zm,Henn:2010bk}. For example, 
products like $\cO(m^2) \times \log m^2$ are evanescent as $m \to 0$, whereas $\cO(\epsilon) \times \epsilon^{-1}$
terms in dimensional regularization must be kept when the regulator is sent to zero. 
Since the IR divergences of the amplitudes produce $\log^{i} m^2 $ behavior, but no polynomial divergences,
% such as e.g. $1/m^2$,
there will be no cross terms between different loop orders.
Of course, individual integrals may diverge linearly in $m^2$ as the mass is taken to zero,
and in this case care is required when they are multiplied by $\cO(m^2)$ terms, see e.g. \cite{Henn:2010ir,Henn:2010kb,Drummond:2010mb}.
Some further aspects of integrals in this regularization are reviewed in section \ref{sect-integrals}. 
\\

A major conceptual motivation for considering the above regulator
is that the string theory dual of $\cN=4$ SYM suggests that it is well adapted to the dual conformal symmetry.
Indeed, the above is inspired by the string theory setup of \cite{Alday:2007hr} (see also \cite{Kawai:2007eg,McGreevy:2008zy,Schabinger:2008ah}.) 
In terms of the latter, the Higgs mass corresponds to the inverse radial coordinate in the AdS${}_{5}$ space, see Fig.~\ref{fone}(a),
or equivalently to the radial coordinate in a dual AdS${}_{5}$ space,
that is obtained by a T-duality transformation. The isometries of the latter suggest a (dual) conformal
symmetry for the scattering amplitudes. The (non-trivial) isometry transformations read,
\begin{align}\label{adsk}
\hat{K}^{\mu} &= \sum_{i=1}^{n} \left[ 2 x_{i}^\mu x_{j}^\mu \frac{\partial}{\partial x^\nu_{i}} - x_{i}^2 \frac{\partial}{\partial x_{i \mu}} + 2 x_{i}^{\mu} m_{i} \frac{\partial}{\partial m_{i}} - m_{i}^2 \frac{\partial}{\partial x_{i \mu}} \right] \,, \\
&= {K}^{\mu} +  \sum_{i=1}^{n} \left[  2 x_{i}^{\mu} m_{i} \frac{\partial}{\partial m_{i}} - m_{i}^2 \frac{\partial}{\partial x_{i \mu}} \right] \,, \nonumber
\end{align}
where we used Poincar\'{e} coordinates $\{x_{i}^{\mu},m_{i} \}$ to parametrize AdS${}_{5}$, 
denoting the radial coordinate by $m$.
For the $m=0$ we recover the standard form of conformal transformations, equation (\ref{kboost}), 
that we used in the previous section.
As we have already seen, the amplitudes are infrared divergent in this case, and therefore
the discussion of the symmetries is only formal. 
On the other hand, for non-zero mass $m$, 
the amplitudes are infrared finite and we have the realization (\ref{adsk}) of the (dual) conformal symmetry.
When using (\ref{adsk}) it is crucial that we have one parameter $m_{i}$ for each dual coordinate $x^{\mu}_{i}$.
This can be achieved by refining the above setup by breaking the gauge group further to $U(N) \times U(1)^M$,
thereby introducing several Higgs masses. 
Given the AdS considerations above, we expect that the scattering amplitudes defined in this way should have an exact dual conformal symmetry, i.e.
\begin{equation}\label{exact-dcs}
\hat{K}^{\mu} M_{n} = 0\,.
\end{equation}
Note that the transformations (\ref{adsk}) also change the value of the $m_{i}$.
In fact, the mass can be thought of as a fifth component of the dual coordinates $x^{\mu}_{i}$.
This means that we should think of the $m_{i}$ as parameters, just like the kinematical
variables of the scattering process.\\

In order to carry out calculations in the field theory, it is important to have an action that
corresponds to the spontaneous symmetry breaking $U(N+M) \to U(N) \times U(1)^{M}$ discussed 
above. The latter and the corresponding Feynman rules were worked out in \cite{Alday:2009zm},
starting from the component action 
\begin{equation}
{\hat S}^{U(N+M)}_{\cN=4}= \int d^{4}x\, \Tr \Bigl (
-\ft{1}{4}\, {\hat F}_{\mu\nu}^{2} -\ft{1}{2}(D_{\mu}{\hat \Phi}_{I})^{2} + \ft{g^2}{4}\, [{\hat \Phi}_{I},{\hat\Phi}_{J}]^{2} +\ft{i}{2}\,\hat{\overline{\Psi}}\,\Gamma^{\mu}\,D_{\mu}\hat{\Psi} +\ft{g}{2}\hat{\overline{\Psi}}\,\Gamma^{I}[{\hat \Phi_{I}},\hat\Psi]\, \Bigr ) \, ,
\end{equation}
where $D_{\mu}=\partial_{\mu} -i g [A_{\mu}, \cdot ]$. 
All fields are hermitian matrices, which we decompose into blocks as
\begin{align}
{\hat A}_{\mu}&=
\begin{pmatrix} % or pmatrix or bmatrix or Bmatrix or ...
      (A_{\mu})_{ab} &  (A_{\mu})_{aj} \\
       (A_{\mu})_{ia} &  (A_{\mu})_{ij} \\
   \end{pmatrix} \, ,\quad
 {\hat\Phi}_{I}=
   \begin{pmatrix} % or pmatrix or bmatrix or Bmatrix or ...
      (\Phi_{I})_{ab} &  (\Phi_{I})_{aj} \\
       (\Phi_{I})_{ia} &  \delta_{I9}\, \ft{m_{i}}{g}\,\delta_{ij}+(\Phi_{I})_{ij} \\
   \end{pmatrix} \, ,\quad
 {\hat\Psi}=
   \begin{pmatrix} % or pmatrix or bmatrix or Bmatrix or ...
      (\Psi)_{ab} &  (\Psi)_{aj} \\
       (\Psi)_{ia} &  (\Psi)_{ij} \\
   \end{pmatrix} \, , 
\end{align}
where $ a,b=1,\ldots, N\, ,i,j=N+1,\ldots, N+M$, and 
we have given a vacuum expectation value $(\vev{\Phi}_{9})_{ij} = \delta_{ij} m_{i}/g$
to the scalars 
%${\hat\Phi}_{I}=\delta_{I9}\, \vev{\Phi_{9}}+\Phi_{I}$
in the $I=9$ direction.
The shift leads to new quadratic and cubic vertices. The former lead to several types of fields. We 
have the `light' fields $\cO_{ij}$ ($i\neq j$) with masses $(m_i-m_j)$, where $\cO$ denotes
a generic field $\{A_{\mu}, \Phi_{I},\Psi\}$, and the heavy fields $\cO_{ia}$ of mass
$m_i$. The $\cO_{ab}$ and $\cO_{ii}$ remain
massless. (In the simplest case $m_{i}=m_{j}$ the `light' fields become massless.)
Moreover, there are new cubic vertices between scalars, and gluons and scalars proportional to $m_{i}$.\\

Let us now see how the exact dual conformal symmetry appears in practice.
A one-loop calculation starting from the action above showed that one obtains the following one-loop four-point amplitude,
\begin{align}
M_{4} = 1 - \frac{a}{2}  I_{4}^{(1)}(s,t,m_{i}) + \cO(a^2)\,,
\end{align}
where
\begin{align}
 I_{4}^{(1)} = \int \frac{d^{4}x_{0}}{i \pi^2} \frac{ (x_{13}^2 + (m_1 - m_3 )^2 ) (x_{24}^2 + (m_2 - m_4 )^2 ) }{  \prod_{i=1}^{4} (x_{0i}^2 + m_i^2 )}
\end{align}
One can now easily check that $I_{4}^{(1)}$ is invariant under the {\it extended} dual conformal
transformations $\hat{K}^{\mu}$.
As before, this is easiest done by noting manifest four-dimensional Poincar\'e symmetry
and applying (dual) conformal inversions to $I_{4}^{(1)}$, where the masses $m_{i}$ are
treated as higher-dimensional components of the dual coordinates $x_{i}^{\mu}$ \cite{Alday:2009zm}.
Infinitesimally, invariance can be expressed as
\begin{align}\label{exact-dcs-invariance}
\hat{K}^{\mu} M_{4}  = 0.
\end{align}
Note that $\mu=0,1,2,3$, i.e. we still have an $SO(4,2)$ symmetry as in the massless case, 
only the representation of the symmetry has changed.
In general, equation (\ref{exact-dcs}) simply implies that $M_{n}$, which a priori is a function of the $m_{i}$ and $p_{i} \cdot p_{j}$,
depends on a restricted set of variables only. E.g. in the four-point case
we have
\begin{align}
M_{4}(m_1, m_2, m_3, m_4 ,x_{13}^2, x_{24}^2 ) = M_{4}(u,v) \,,
\end{align}
where
\begin{align}\label{uv-variables}
 u =  \frac{m_1 m_3}{ x_{13}^2 + (m_1 - m_3 )^2 }\,,\qquad  v= \frac{m_2 m_4 }{x_{24}^2 + (m_2 - m_4 )^2} \,.
\end{align}
In the above example the fact that the integral $I_{4}^{(1)}$ depends on $u$ and $v$ only
can also be seen directly in the Feynman parametrisation of this integral, by rescaling the 
Feynman parameters for each propagator $1/(x_{0i}^2 + m_{i}^2)$ by $m_{i}$.\\

A natural conjecture is that at a given loop level, the amplitude can be written as a linear
combination of integrals $I_{\sigma}$ invariant under {\it extended} dual conformal symmetry \cite{Henn:2010bk,Henn:2010ir},
\begin{align}\label{edci-ansatz}
M_{n}^{(L)} = \sum_{\sigma} \, c_{\sigma} \, I_{\sigma} \,,
\end{align}
where
\begin{equation}\label{edci-condition}
\hat{K}^{\mu} I_{\sigma} = 0\,,
\end{equation}
i.e. the $I_{\sigma}$ are invariant under the extended dual conformal symmetry,
and the $c_{\sigma}$ are rational coefficients (e.g. numbers in the MHV case or 
in general dual conformal invariants similar to those that appear in the tree-level amplitude \cite{Drummond:2008cr}.)
This is exactly what many authors suspected,
using the notion of ``pseudoconformal'' integrals.
The latter can now be replaced by the concise definition (\ref{edci-condition}).
As was already explained, equation (\ref{edci-ansatz}) has important practical consequences,
e.g. when computing loop amplitudes through the unitarity method.\footnote{Given equation (\ref{edci-ansatz}) for amplitudes on the Coulomb branch, one may wonder what its
consequences are for amplitudes at the origin of the Coulomb branch and for $D \approx 4$. We caution the
reader that switching between IR regularizations at intermediate steps of a calculation is in general very subtle, especially at higher loop orders.
 }\\

We have seen above that exact dual conformal symmetry reduces the number of variables
that a function can depend on.
It is important to realize that it is a stronger constraint to require that such a function should
come from a loop integral, 
i.e. that is built from propagators that are integrated over
space-time.
For example, in Feynman gauge the propagator denominators are always $1/p^2$ or $1/(p^2+m^2)$,
and at $L$ loops we have $L$-fold loop integrals built from such propagators and possibly numerator 
factors as a result of the numerator algebra.
This is important in two respects. Firstly, for a given scattering amplitude, one can classify the
loop integrals having this property, which are naturally much fewer than the set of generic loop integrals.
Secondly, the fact that the functions we are dealing with come from loop integrals means that
we can use properties of the latter such as their analytic structure, unitarity cuts, etc. \cite{Eden}, to infer properties
of the functions (see also the comments in the next section.)\\

The interpretation of the masses as components of higher-dimensional momenta
motivated several groups to investigate dual conformal symmetry in higher dimensions.
It was shown that tree-level (super)amplitudes in six dimensions \cite{Bern:2010qa,Dennen:2010dh} 
and ten dimensions \cite{CaronHuot:2010rj} have a dual conformal symmetry.
In turn, since the higher-dimensional amplitudes can be interpreted as the massive four-dimensional Coulomb branch amplitudes of \cite{Alday:2009zm},
this proves that the latter are indeed dual conformally invariant at tree level.\\

This also has important consequences for loop-level amplitudes on the Coulomb branch
of $\cN=4$ SYM, as it essentially proves the conjectures made in \cite{Alday:2009zm}.
Previous evidence in support of these had come from \cite{Henn:2010bk,Boels:2010mj,Henn:2010ir,Bern:2010qa}.
It was shown in \cite{Dennen:2010dh} that all unitarity cuts of planar loop
amplitudes in that theory have the (extended) dual conformal symmetry. This proves the (extended) dual conformal 
symmetry of loop amplitudes in $\cN=4$ SYM, up to potential terms not detected by any unitarity
cuts.
Similarly one can argue that in theories where tree amplitudes determine the loop
integrand, e.g. through recursion relations \cite{CaronHuot:2010zt,ArkaniHamed:2010kv,Boels:2010nw}, the latter should inherit the (extended) dual conformal symmetry
from the trees \cite{CaronHuot:2010rj}. 
\\

As was explained above, the restrictions imposed above from (extended) dual conformal
symmetry on the loop amplitudes are very useful when determining the loop integrand through
the (generalized) unitarity method. Recently, it was realized that the BCFW idea \cite{Britto:2004ap,Britto:2005fq} of determining
tree-level amplitudes from their factorization channels can also be applied to planar loop integrands \cite{CaronHuot:2010zt,ArkaniHamed:2010kv,Boels:2010nw}.
The loop integrand of a given amplitude can then be iteratively determined starting from tree amplitudes
in the forward limit (see also \cite{Bierenbaum:2010xg} and references therein.) In practice this works extremely well for computing the loop integrand in four dimensions,
since the corresponding tree-level amplitudes are known \cite{Drummond:2008cr}. In order to obtain an integrand that
can be safely integrated one should in principle determine e.g. the $D$-dimensional loop integrand (for
dimensional regularization), or the integrand on the Coulomb branch. Given the four-dimensional integrand,
the extended dual conformal symmetry provides useful guidance for how to ``translate'' the latter to the Coulomb branch
integrand, and it is argued that this should give the correct integrand, up to $\cO(m^2)$ corrections \cite{ArkaniHamed:2010kv}.
We note another interesting recent approach to loop integrands that is based on a connection to correlation functions \cite{Alday:2010zy}.

\section{Properties of the loop integrals}
\label{sect-integrals}

Here we make a number of comments on properties of the loop integrals,
with a special focus on the mass regulator. We comment on their evaluation
using Mellin-Barnes methods, their properties in the Regge limit, and
review a type of dual conformal integrals with special numerators. 
The latter integrals satisfy simple differential equations. We comment
on a possible relation to conventional conformal symmetry.\\

A state-of-the-art tool for evaluating loop integrals is the
Mellin-Barnes (MB) method \cite{smirnov2006feynman}, where one trades Feynman
parameter integrals for contour integrals by (repeatedly) using the identity
\begin{eqnarray}
(X+Y)^{-\lambda} = \frac{ 1}{\Gamma(\lambda)}  \oint_{\beta - i \infty}^{\beta+i \infty} \frac{dz}{2 \pi i} \frac{Y^z}{X^{\lambda+z}} \Gamma(-z) \Gamma(z+ \lambda) \,,
\end{eqnarray}
with $\beta <0$.
This approach works well
for the massless as well as for the massive case. Experience shows \cite{smirnov2006feynman} that introducing
the Mellin-Barnes parameters loop by loop is a good strategy. 
Moreover, in the present case, one can often perform all 
manipulations while staying in $D=4$ dimensions.
This should be done whenever possible to obtain a low-dimensional
Mellin-Barnes representation. It is interesting to note that
starting from the four-loop level, the massive MB representations tend
to involve fewer parameters as compared to the dimensional regularization case.
A very detailed derivation of the MB representations for the massive three-loop
four-point integrals is given in appendix A of \cite{Henn:2010bk}.\\

\begin{figure}
 \psfrag{A}[cc][cc]{$(a)$}
\psfrag{B}[cc][cc]{$(b)$}
\psfrag{s}[cc][cc]{$s$}
\psfrag{t}[cc][cc]{$t$}
\psfrag{p1}[cc][cc]{$p_{1}$}
\psfrag{p2}[cc][cc]{$p_{2}$}
\psfrag{p3}[cc][cc]{$p_{3}$}
\psfrag{p4}[cc][cc]{$p_{4}$}
\psfrag{dots}[cc][cc]{$\dots$}
\psfrag{regge}[cc][cc]{$s \gg t$}
\psfrag{conj}[cc][cc]{}
\psfrag{ILa}[cc][cc]{$I_{L\,a}$}
\psfrag{ILH}[cc][cc]{$I_{L\,H}$}
\psfrag{LLandNLL}[cc][cc]{LL and NLL }
\psfrag{NLL}[cc][cc]{NLL}
 \centerline{
 {\epsfxsize15cm  \epsfbox{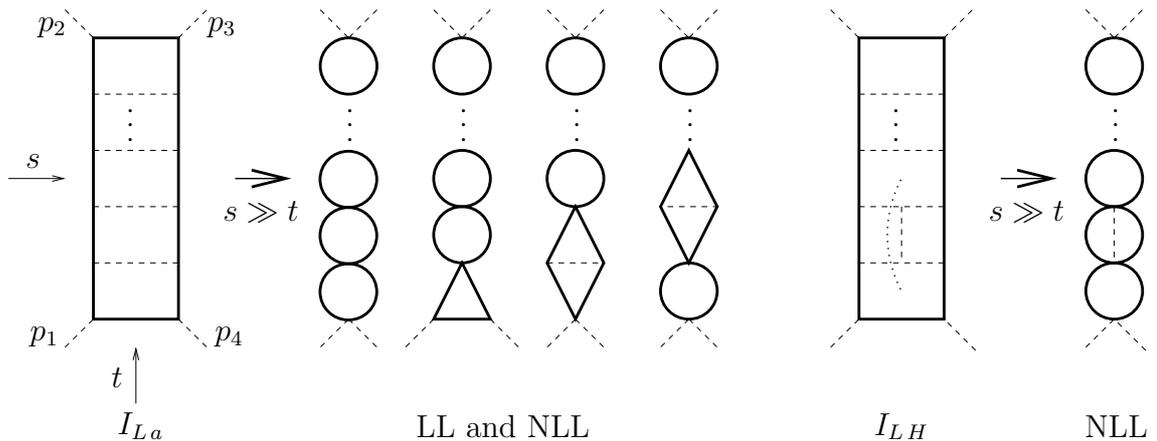}}
}
\caption{\small
Factorization of the leading-log and next-to-leading-log contributions 
to the Regge limit $s \gg t$ 
of the $L$-loop vertical ladder integral $I_{L\,a}(v,u)$
into simpler integrals. 
Factorization of the NLL contribution of the
vertical ladder integral with H-shaped insertion $I_{L\,H}$ \cite{Henn:2010bk,Henn:2010ir}. 
The dotted line indicates a loop-momentum-dependent numerator. Picture from \cite{Henn:2010ir}.}
\label{fig-regge1}
\end{figure}

One advantage of the Higgs setup is that it is natural to consider the amplitudes and 
integrals for finite values
of $m^2$. In this spirit, one can consider the Regge limit, e.g. $s \gg t, m^2$ in
the four-particle case. In \cite{Henn:2010bk,Henn:2010ir} it was shown that the integrals contributing to
the amplitudes behave very nicely in this limit. One can show that to all loop orders,
the leading log (LL) and next-to-leading log (NLL) contribution to the Regge limit 
is given by the two infinite classes of integrals shown in Fig.~\ref{fig-regge1}.\\

For example, the contribution of $I_{L\,H}$ to NLL accuracy is given by (we use the notation $u=s/m^2$ and $v=t/m^2$)
\begin{equation}
\label{LH}
 I_{L\,H} =   \frac{(-1)^{L-1} }{(L-1)!}  \log^{L-1} u \times
 K(v)^{L-2} \times K'(v)
   \quad  +\quad  \cO(\log^{L-2} u ) \,,
\end{equation}
where $K(v)$ and $K'(v)$ correspond to the two-dimensional bubble
and two-loop bubble diagrams shown in fig.~\ref{fig-regge1}  (see
ref.~\cite{Henn:2010bk} for further discussion).
Taking $v$ small, we have
\begin{eqnarray}
K(v) &=& -2 \log v + O(v)\,, 
\nonumber\\
K'(v) &=& -\frac{4}{3} \log^{3} v -  \frac{4}{3} \pi^2 \log v + \cO(v) \,,
\end{eqnarray}
which, combined with the result for $I_{L\,a}(v,u)$ gives the correct Regge behavior
at LL and NLL \cite{Henn:2010ir,Henn:2010ir}.
Notice that the fact that the $I_{L\,H}$ starts contributing at NLL and not NNLL is
possible only thanks to its non-trivial numerator factor, whose presence in turn
is required by dual conformal invariance.
It is interesting to note that such non-trivial (loop-momentum-dependent) numerator
factors are also important when discussing UV properties of scattering amplitudes
in $\cN=4$ SYM and $\cN=8$ supergravity \cite{Bern:2009kd}.\\

Dual conformal symmetry can also lead to interesting insights about the 
asymptotic behavior of integrals/amplitudes in certain limits. Recall that
the integrals can depend on the masses only in specific combinations with
the kinematical variables, see e.g. (\ref{uv-variables}) in the four-point case.
This implies that certain small mass limits are equivalent to Regge limits.
See \cite{Henn:2010bk,Henn:2010ir} for more details.\\

Recently it has become apparent that it is particularly advantageous
to introduce dual conformal integrals with certain non-trivial numerator factors 
 \cite{ArkaniHamed:2010kv}.
A guiding principle in defining these numerator factors are (potential) infrared divergences.
The latter can arise from specific integration regions where loop propagators go on shell.
If the appropriately defined numerator factors vanish in those regions they will
soften the infrared divergences of the integral, or even make the integral finite.
Let us give a simple example of the latter type. Consider the following 
pentagon integral with two off-shell and three on-shell legs.
In dual coordinates, it is defined by
\begin{eqnarray}\label{7ptpent}
{
 \psfrag{x1}[cc][cc]{$x_{1}$}
 \psfrag{x3}[cc][cc]{$x_{3}$}
 \psfrag{x4}[cc][cc]{$x_{4}$}
 \psfrag{x5}[cc][cc]{$x_{5}$}
 \psfrag{x6}[cc][cc]{$x_{6}$}
 \psfrag{xr}[cc][cc]{}
\parbox[c]{30mm}{\includegraphics[height = 25mm]{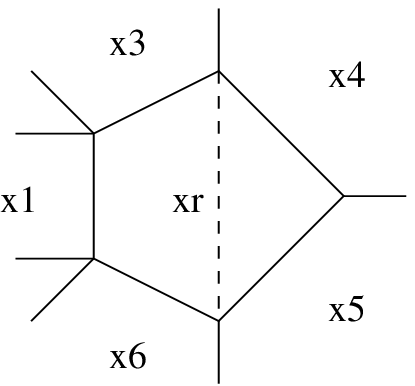}}
}
:= \frac{x_{14}^2 x_{15}^2 x_{36}^2}{x_{1a}^2}\, \int \frac{d^{4}x_{r}}{i \pi^2} \, \frac{x_{ar}^2}{x_{1r}^2 x_{3r}^2 x_{4r}^2 x_{5r}^2 x_{6r}^2 } \,,
\end{eqnarray}
where $x_{34}^2 = x_{45}^2 = x_{56}^2 = 0$ and where the ``magic point'' $x_{a}^{\mu}$, which is denoted by a dashed line on the l.h.s. of (\ref{7ptpent}), is
defined as one of the solutions to the four-cut condition
\begin{eqnarray}\label{cond-4cut}
x_{3a}^2 = x_{4a}^2 = x_{5a}^2 = x_{6a}^2 = 0 \,.
\end{eqnarray}
Hence we see that the numerator factor vanishes in the regions that would otherwise
produce infrared divergences, and the integral is finite.
The above example is sufficiently simple that we will not need to write out
the explicit solution to (\ref{cond-4cut}), and we can compute it using e.g. Feynman parameters.
(In general the explicit definition of the numerator factors
can be written very conveniently using momentum twistor variables \cite{Hodges:2009hk}.
At the loop level, the latter are ideally used in combination with the above mass regularization, as they are
intrinsically four-dimensional. See \cite{Mason:2010pg,Hodges:2010kq} for more details.)
Being dual conformally invariant, the answer is a function of the cross-ratios $u_1 = (x_{13}^2 x_{46}^2)/(x_{14}^2 x_{36}^2)$
and $u_2 = (x_{16}^2 x_{35}^2)/(x_{15}^2 x_{36}^2)$. Multiplying for convenience by $(1- u_1 - u_2 ) $, 
one obtains the simple formula
\begin{eqnarray}
{\Psi}^{(1)}(u_1 , u_2 ) :=  (1- u_1 - u_2 ) \times {\rm Eq.}(\ref{7ptpent}) =  \log u_1 \log u_2 + {\rm Li}_{2}(1- u_1 ) + {\rm Li}_{2}(1- u_2 ) - \zeta_{2}  \,.
\end{eqnarray}
It is important to note that this integral is related to standard integrals by simple integral
reduction identities \cite{Drummond:2010mb,ArkaniHamed:2010kv}.
In the present case, one could represent the pentagon integral above
by a linear combination of five (IR-divergent) one-mass box integrals.\\

The idea is then that integrals of the type discussed above can be used, 
thanks to the numerator identities, to write loop amplitudes in a simpler form.
For example, a good strategy could be to trade the most complicated integrals
in a given calculation (say, double pentagon integrals) for the integrals discussed
here, and simpler integrals.
For example, when applying these ideas to the $n$-point two-loop MHV amplitudes in $\cN=4$ super Yang-Mills,
one obtains very compact expressions \cite{ArkaniHamed:2010kv,Drummond:2010mb}. In fact, only one integral topology 
appears, with different arrangements of external legs. In total, only
$36$ distinct integrals are needed to fully describe the two-loop MHV amplitudes
with arbitrary number of external legs. Many of these integrals are inter-related by soft
limits.\\

Moreover, it was found in \cite{Drummond:2010mb} that the new integrals, when evaluated, lead to rather
simple functions, just as in the one-loop example above.
This allowed e.g. the analytical computation of the six-point remainder
function in kinematical limits \cite{Drummond:2010mb}. This is the first time that this was achieved directly from
the loop integrals (previous analytical results were available from Wilson loop calculations \cite{DelDuca:2010zg}.)
The simplicity of the integrals is explained (in part) by the fact that they satisfy simple differential equations \cite{Drummond:2010cz}.
For example, for the pentagon example discussed above one can show that
\begin{eqnarray}
u_2 \partial_{u_2 }  u_1 \partial_{u_1 }  {\Psi}^{(1)}(u_1 , u_2 ) = 1\,.
\end{eqnarray}
It was found in \cite{Drummond:2010cz} that the integrals relevant for planar MHV amplitudes 
in $\cN=4$ SYM satisfy similar differential equations, which relate in general $L$-loop to $(L-1)$-loop integrals.
Apart from helping in finding analytical solutions, see \cite{Drummond:2010cz} for several non-trivial examples
at the two-loop level, the simple nature of the equations also suggest that their solutions cannot have an
arbitrarily complicated structure.
\\

\begin{figure}
 \psfrag{a}[cc][cc]{$(a)$}
\psfrag{b}[cc][cc]{$(b)$}
\psfrag{p1}[cc][cc]{$p_{1}$}
\psfrag{p2}[cc][cc]{$p_{2}$}
\psfrag{p3}[cc][cc]{$p_{3}$}
\psfrag{p4}[cc][cc]{$p_{4}$}
\psfrag{x1}[cc][cc]{$x_{1}$}
\psfrag{x3}[cc][cc]{$x_{3}$}
\psfrag{x5}[cc][cc]{$x_{5}$}
\psfrag{x7}[cc][cc]{$x_{7}$}
\psfrag{y1}[cc][cc]{$y_{1}$}
\psfrag{y2}[cc][cc]{$y_{2}$}
\psfrag{y3}[cc][cc]{$y_{3}$}
\psfrag{y4}[cc][cc]{$y_{4}$}
\psfrag{y5}[cc][cc]{$y_{5}$}
\psfrag{y6}[cc][cc]{$y_{6}$}
\psfrag{y7}[cc][cc]{$y_{7}$}
\psfrag{y8}[cc][cc]{$y_{8}$}
 \centerline{
 {\epsfxsize10cm  \epsfbox{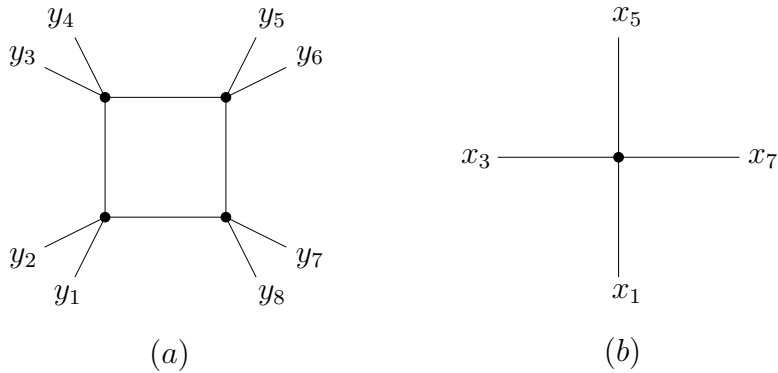}}
}
\caption{\small
The integral of equation (\ref{eq-yangian-int}) in momentum space (a) and in dual notation (b).
The position space variables $y_{i}^{\mu}$ are related to the momenta $p^{\mu}_{i}$ by Fourier transform, 
the dual coordinates $x^{\mu}$ are defined by equation (\ref{dualcoordinates}).
The original and the dual diagram are both built from quartic vertices only, so that the integral has both
a conformal as well as a dual conformal symmetry.
}
\label{fig-Yangian-int}
\end{figure}
It would be interesting to understand to what extent this is a manifestation of the underlying Yangian
symmetry \cite{Drummond:2009fd,ArkaniHamed:2010kv} of scattering amplitudes in planar $\cN=4$ SYM. 
As we have seen, the dual conformal symmetry is under full control at loop level thanks to the
mass regulator, so the question is whether one can put to use the underlying conformal symmetry
of the massless scattering amplitudes. 
As a motivation, we note that Yangian symmetric quantities at loop level do exist.
Here we use Yangian symmetry in a somewhat loose way, meaning conformal and dual conformal symmetry.
Consider for example the following integral, see Fig~\ref{fig-Yangian-int}(a), which could appear in an eight-particle scattering amplitude
in scalar $\phi^4$ theory in four dimensions,\footnote{Related discussions with J.~Drummond and J.~Plefka are gratefully acknowledged.}
\begin{equation}\label{eq-yangian-int}
\int \frac{d^{4}x_{r}}{i \pi^2} \, \frac{1}{x_{r1}^2 x_{r3}^2 x_{r5}^2 x_{r7}^2} \,.
\end{equation}
Since two on-shell legs enter each corner of the box
integral, the corresponding momenta, e.g. $p^{\mu}_{1} + p^{\mu}_{2}$ are off-shell, i.e. $(p_{1}+p_{2})^2= x_{13}^2 \neq 0$, and the integral is
finite in four dimensions. By the analysis of section \ref{sect-historical}, it is also dual conformally covariant,
as can be seen from its dual graph in Fig.~\ref{fig-Yangian-int}(b).
Moreover, because of its origin as a finite graph in $\phi^4 $ theory, it is also conformally
invariant. This is easiest seen in position space.
The conformal symmetry leads to second-order homogeneous differential equations for the integral.
In this example the integral is effectively off-shell, and there are no IR divergences at all, whereas in $\cN=4$ SYM one would
first have to separate IR-divergent and IR-finite pieces in a convenient way. Depending on how
this is done, it is plausible that one could find homogenous or inhomogeneous differential equations
as a manifestation of the underlying symmetry. 
In this spirit it would be interesting if the differential equations found
in \cite{Drummond:2010cz} could be related to, or understood more systematically in terms of the underlying conformal symmetry
of $\cN=4$ SYM.

\section{Conclusion}

In this article we have reviewed the current status of dual conformal symmetry 
at loop level in planar $\cN=4$ SYM. The best way to understand this symmetry at loop
level is on the Coulomb branch of $\cN=4$ SYM and by using a representation
that is suggested by the isometries of AdS${}_{5}$. The Coulomb branch amplitudes have
an exact dual conformal symmetry. The latter leads to powerful constraints for
the loop integrand of the scattering amplitudes.\\

New recursion relations for loop integrands provide a powerful practical tool for
determining the latter. The {\it four-dimensional} loop integrand can be easily obtained, 
and dual conformal symmetry helps convert the latter to the correct
Coulomb branch integrand, up to $\cO(m^2)$ corrections \cite{ArkaniHamed:2010kv}. 
Given this, the main task for solving planar $\cN=4$ SYM lies in the evaluation of the
(dual conformal) loop {\it integrals}. 
Here the formulation in terms of momentum twistor integrals, where necessary in combination with
the mass regulator, seems very promising.\\

The ultimate goal is to obtain results that can interpolate between weak and strong coupling \cite{JPAstrong}.
In fact there are integrals closely related to the ones discussed here, for which all-loop results are available, 
and where a resummation is possible \cite{Broadhurst:2010ds}. The differential equations found in \cite{Drummond:2010cz} provide 
hope that this may be possible for the integrals directly relevant to $\cN=4$ SYM as well.

\section{Acknowledgements}
We thank Z.~Bern for correspondence and J.~Plefka and R.~Roiban for a careful reading of the manuscript and for useful comments.

\bibliographystyle{nb.bst}
\bibliography{dcsloop}

\end{document}